\documentstyle[12pt,epsf]{article}

\begin{document}
\begin{flushright}
ITP-SB-95-30 \\
hep-ph/9508358
\end{flushright}
\smallskip

\vbox{\vskip 1.0 true in}

\centerline{\Large \bf Soft Gluon Resummation\footnote{Based on a talk
presented at the 10{\it th} Topical Workshop on Proton-Antiproton
Collider Physics, Fermilab, May 9-13, 1995}}
\bigskip

\centerline{\large George Sterman}

\smallskip
\centerline{\it Institute for Theoretical Physics}
\centerline{State University of New York, Stony Brook, NY 11794-3840}
\vbox{\vskip 1.0 true in}

\begin{abstract}
I discuss some of the basic techniques and results of soft
gluon resummation, and their applications to scattering
at collider energies.
\end{abstract}
\newpage

\section{Varieties of Factorization and Evolution}

With the extraordinary data from the Tevatron of the last
runs, our knowledge of large momentum transfer processes has taken
a qualitative step forward.  The demands on theory are
now much more stringent, and afford tests of
QCD concepts that
were not previously practical.  In particular, TeV
energies allow for many two-scale processes where both
scales are much larger than a GeV, and hence may be amenable
to perturbative treatment.  For instance, $\alpha_s(t)\; \ln(s/t)$
may be a relatively large number
even for $\alpha_s(t)$ small, necessitating a resummation
of powers $(\alpha_s(t)\; \ln^2(s/t))$
and $(\alpha_s(t)\; \ln(s/t))$
to all orders in perturbation theory.

In the following, I shall review the basic techniques
used to study two-scale inclusive cross sections at
low parton density.  For such process factorization
in terms of single-parton densities holds,
\begin{equation}
\sigma_H(p,q) = \sum_i \int dx' {\hat \sigma}_i(x'p,q)\; \phi_{i/H}(x')\, ,
\label{factor}
\end{equation}
with ${\hat \sigma}_i$ a short-distance partonic cross section
and $\phi_{i/H}$ the distribution of parton $i$ in hadron $H$.

I will begin with a review of
three basic evolution equations, which govern
the behavior of parton distributions in three limits
of particular physical interest, and which are commonly
referred to as the DGLAP, BFKL and Sudakov equations.
Of these, the DGLAP is the best-studied, and
controls the standard evolution of parton distributions
in $Q^2$ for values of $x$ for which neither $\alpha_s\ln x$ nor
$\alpha_s \ln(1-x)$ is a large number.
The BFKL equation governs the behavior of
parton distributions at small $x$ and fixed $Q^2$ and of
amplitudes near the forward direction.  Finally, the
Sudakov evolution equation describes the behavior
of distributions and amplitudes in the elastic limit
($x\rightarrow 1$ for deep-inelastic scattering).  These
limits are illustrated schematically in Fig.\ 1.
\begin{figure}[ht]
%
\centerline{\epsffile{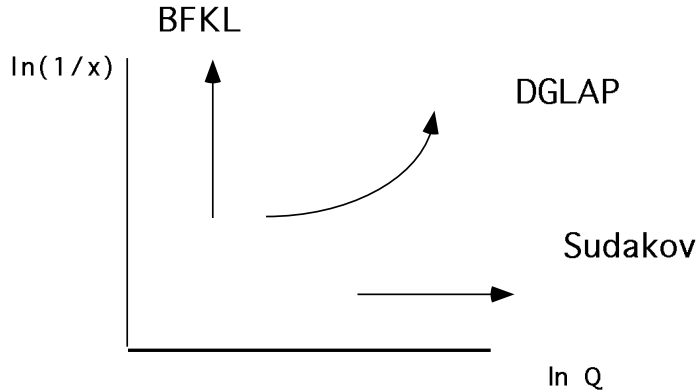}}
\caption{Actions of evolution equations.}
\label{fig1}
\end{figure}

Each of these evolution equations is associated with the ladder structure
shown in Fig.\ 2, in which parton $j$, with momentum $p$
``evolves" into parton $i$, of momentum $xp$, in the process
emitting $N$ quanta.  In perturbation theory, the
distribution of parton $i$ in parton $j$ may be thought
of as a sum over $N$,
\begin{equation}
\phi_{i/j}(x,q_T) \sim \sum_N \phi_{i/j}^{(N)}\, .
\label{laddersum}
\end{equation}
\begin{figure}[ht]
\centerline{\epsffile{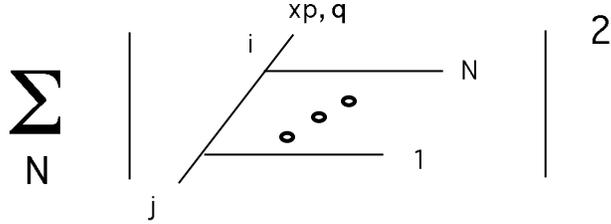}}
\caption{General ladder structure.}
\label{fig2}
\end{figure}

\subsection{DGLAP}

The basic evolution equation for perturbative QCD, originated by
 Dokshitser, Gribov, Lipatov, Altarelli and Parisi
\cite{DGLAP}, has its origin in the description of deeply inelastic
scattering (DIS).  The factorization theorem for any DIS
structure function $F$ in hadron $H$ is
\begin{equation}
F^{(H)}(x,Q^2) = \int_x^1 {dx'\over x'}\; C_{j}(x/x')\; \phi_{j/H}(x,Q^2)\, ,
\label{DISfact}
\end{equation}
with perturbative short-distance functions $C_i$ and nonperturbative
parton distribution(s) $\phi$.  Although $\phi$ is
not perturbatively calculable, its $Q^2$ dependence is, in terms
of kernels, or splitting functions, $P_{ij}$, through the DGLAP evolution
equation
\begin{equation}
{ d\over d \ln Q^2}\phi_{i/H}(x,Q^2)
=
\sum_{j=f,{\bar f},G} \int_x^1 {d \xi \over \xi}\; P_{ij}(x/\xi)\;
\phi_{j/H}(\xi,Q^2)\, .
\label{dglap}
\end{equation}
The universality of the distributions allow us to
derive predictions for hadronic collisions from parton
distributions derived (primarily) from DIS.  The
$P_{ij}$ themselves are known up to two loops.

In general, the DGLAP equation is applicable when the evolution
is generated by the emission of quanta with strongly ordered
transverse momenta ($k_{1T}\ll k_{2T} \ll \dots \ll k_{NT}$ in
Fig.\ 2, for instance).  The integrals over these
transverse momenta give logarithmic enhancements, which are
organized by the DGLAP equation.
 There are other sources of logarithmic
enhancement, however, already present as factors of $1/x$ in
the splitting function $P_{GG}(x)$.  Such singular behavior produces
logarithms of $x$, even in the absence of large transverse momentum
enhancements.  The BFKL equation, to which we now turn,
summarizes the effects of these logs of $x$.

\subsection{BFKL}

The BFKL equation (Balitskii, Fadin, Kuraev and Lipatov \cite{BFKL})
was developed originally to resum logs of $s$ in the total hadronic
cross section (see below).  We begin our consideration of
it here, however, with a generalization of
DIS factorization (\ref{DISfact}), to a form that links the
distributions in both longitudinal and transverse momenta \cite{ktfactci},
\begin{equation}
F(x,Q^2)=\int_x^1 {dx' \over x'}\; d^2k_T\; C(x/x',{\bf k},Q^2)\, {\cal
F}(x',{\bf k})\, ,
\label{ktfact}
\end{equation}
with $C(\xi,{\bf k},Q^2)$ a new, transverse momentum-dependent
short-distance function and $\cal F$ the corresponding parton
distribution.  In general, the gluon distribution tends to
dominate at low $x$, and we shall restrict our discussion to
${\cal F}_G$.  Eq.\ (\ref{ktfact}) is of particular interest
when $x\rightarrow 0$, so that the mass of the final state
in DIS, $(1-x)Q^2/x$, grows.

The BFKL equation, which describes the behavior of ${\cal F}_G$,
may be written in many forms, one of the least intimidating
of which is
\begin{equation}
{d\over d \ln x}{\cal F}_G(x,{\bf k})
=
{3\alpha_s\over \pi} \int{d^2{\bf k}' \over ({\bf k}'-{\bf k}')^2}\
\left ( K\; *\; {\cal F}_G(x,{\bf k}') \right )\, ,
\label{bfkl}
\end{equation}
with a kernel defined by
\begin{eqnarray}
K\; *\; {\cal F}_G(x,{\bf k}')
&=&
{\cal F}_G(x,{\bf k}') \nonumber \\
&\ & \quad -
\left ( {{\bf k}\cdot{\bf k}'\over 2{{\bf k}'}^2} \right )
{\cal F}_G(x,{\bf k})\, .
\label{bfklkernel}
\end{eqnarray}
An excellent introduction to the BFKL equation may be
found in the recent lectures of Del Duca \cite{DelDuca}.

The BFKL equation is somewhat harder to solve than the DGLAP equation,
but it has many applications, including DIS for small $x$, semihard
processes (minijet, heavy quark, etc.) in hadronic collisions,
and color singlet exchange at $|t|\ll s$.  Although the BFKL
formalism has traditionally remained at leading logarithm in $x$,
recent progress has been reported on generalizations that determine
a two-loop kernel \cite{BFKL2loop}.
In addition, it is possible to develop a generalized evolution
equation that interpolates between the DGLAP and BFKL equations
at leading logarithm \cite{Marches}.
For this purpose, the angle of emission for soft gluons
may be chosen as the primary evolution variable,
reflecting an ordering of angles in sequential emission \cite{Muellangord}.
Control of such ``coherence" effects in soft
QCD radiation \cite{cohere} also serves as an
important ingredient in the construction of
detailed models of QCD final states \cite{herwig}.

\subsection{Sudakov}

The other fundamental limit of evolution is $x\rightarrow 1$.  In DIS,
this corresponds to a low-mass final state with high energy.  This
``elastic" limit is illustrated in Fig.\ 3.  In addition to soft
radiation, a well-collimated jet of mass near $(1-x)Q^2$
emerges from the hard scattering $H$.
\begin{figure}[ht]
\centerline{\epsffile{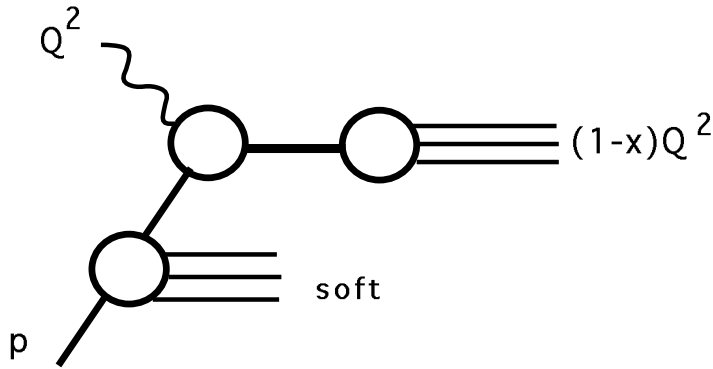}}
\caption{DIS as $x\rightarrow 1$.}
\label{fig3}
\end{figure}
Corresponding to this
physical situation, we may identify a factorized form
for such a cross section,
\begin{equation}
F(x,Q^2)
=
H(Q)\; \int_x^1 dx'\; J((1-x')Q^2,Q)\, \phi(x',Q^2)\, ,
\label{Sudfact}
\end{equation}
with $J$ a universal function representing the hadronization
of the jet (in this form certain non-leading
logarithmic factors are suppressed).  For $(1-x)Q^2\gg \Lambda^2$, $J$ is
computable
in perturbation theory.
Given the factorization (\ref{Sudfact}), the $x$ dependence
of the jet may be determined from the Sudakov evolution equation
\cite{St87},
\begin{eqnarray}
\left ( {\partial \over \partial \ln(1-x)} +
 {1\over 2}\beta(g){\partial \over \partial g}\; \right ) J((&1&-x')Q^2,Q)
= {1\over 2}S(\alpha_s)\; J((1-x')Q^2,Q)
\nonumber \\
&\, &
-{1\over 2}\int_x^1 {d y\over y}\; K_J(1-x/y)\; J((1-y)Q^2,Q)\, ,
\label{Sudevol}
\end{eqnarray}
whose kernel $K_J$ is of the form of a plus distribution,
\begin{equation}
K_J(1-z) = \left [ {\Gamma_J(\alpha_s((1-z)^2Q^2)\over 1-z} \right ]_+\, .
\label{Sudkernel}
\end{equation}
The function $\Gamma_J$ is a universal Sudakov anomalous
dimension \cite{CoSo81}, related to the $\ln n$ dependence of
the standard DGLAP anomalous dimensions $\gamma_n$ \cite{KoTr}.
Thus, $\Gamma_J$ is also known to two loops.
$S$ is a power series in $\alpha_s(Q^2)$.  Sudakov evolution
finds applications to jet event shapes, threshold
corrections and transverse momentum distributions, some of
which we shall review below.

\section{Resummation for Small $x$ and Diffractive Cross Sections}

In this section, I will review a few of the prominent applications of
the BFKL resummation of small-$x$ enhancements, beginning with
DIS.

\subsection{DIS}

For DIS, the object of interest is the $k_T$-dependent
gluon distribution, the solution to eq.\ (\ref{bfkl}).
Eigenfunctions of the derivative may be found by direct
substitution.  The dominant power as $x\rightarrow 0$
specifies the growth of the gluon distribution.  It is \cite{BFKL,DelDuca}
\begin{equation}
{\cal F}_G(x,{\bf k}) \sim x^{-\omega_0}\, ,
\label{bfklsoln}
\end{equation}
times a power of ${\bf k}^2$, where
\begin{equation}
\omega_0={4N\alpha_s \over \pi}\ln 2\, ,
\label{omegao}
\end{equation}
with $N=3$ the number of colors.
Since quark distributions mix with the
gluon distribution, BFKL
evolution suggests that
the structure function of any hadron behaves as
\begin{eqnarray}
F_2(x) &\sim& \sum_i Q_i^2\; x\; \phi_i(x) \nonumber \\
&\rightarrow& x^{-\omega_0}\, ,
\label{ftwopom}
\end{eqnarray}
which diverges as a power.  This behavior is referred to
as that of the ``bare" pomeron.  It is clear that such growth
cannot be supported to arbitrarily small $x$, since
if parton densities grow too large, they
will begin to interfere or ``shadow" each other.
In technical terms, the assumption of low-density
partons, which underlies each of the
evolution equations above, fails for $x$ low
enough \cite{shadow}.

HERA \cite{zeusH1} has seen a growth in DIS structure
functions at small $x$ which is consistent with
the qualitative expectations of BFKL
resummation.  The DGLAP equation, however, based
on $k_T$-ordering, also results in growth for
$x$ small, if not quite so dramatic.  Untangling
the physical content of the HERA data is a subject
of great current interest \cite{smallxtheo}.

\subsection{BFKL in Hadron-hadron Scattering}

The BFKL formalism \cite{BFKL} was originally developed to describe
hadron-hadron scattering in QCD, both the total
cross section and the closely-related Regge limit,
$t$ fixed, $s\rightarrow \infty$.
Its basic consequences for inclusive hadron-hadron scattering near
the forward direction may by summarized by the following
``translation" from DIS notation,
\begin{eqnarray}
M^2_{\rm had} = {1-x \over x}Q^2 &\rightarrow& s
\nonumber \\
        x^{-\omega_0} &\rightarrow& s^{\omega_0}
\nonumber \\
             F(x) &\rightarrow& \sigma_{\rm  tot} \sim s^{\omega_0}\, .
\label{distohh}
\end{eqnarray}
Note in particular that the total cross section grows as a
power, corresponding to the $x\rightarrow 0$ behavior of
$F_2(x)$ in DIS.  Again, a rise in the total cross section
for hadron-hadron scattering has long been seen at high energy,
but an uninterrupted
 power-law rise would violate unitarity, as embodied
in the Froissart bound.  The ``bare" pomeron identified above
therefore cannot be the final answer.  I shall return to
recent progress on this question below.

It is of some interest to sketch the momentum-space
configurations that give rise to the BFKL pomeron in
hadron-hadron scattering.  These are illustrated
by the ladder diagrams in Fig.\ 4,
where the vertices $C_i$ summarize contributions not
only from the diagrams explicitly shown, but also from
non-ladder diagrams
of the same order that are important in gauge theories \cite{BFKL}.
\begin{figure}[ht]
\centerline{\epsffile{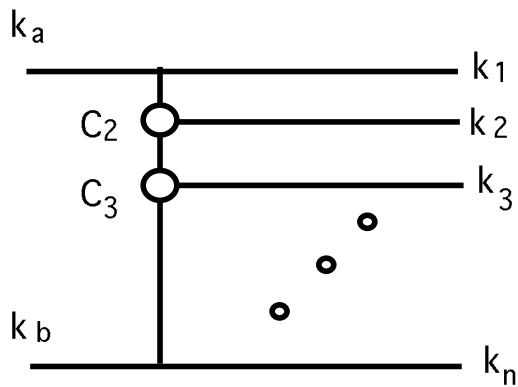}}
\caption{BFKL ladders.}
\label{fig4}
\end{figure}
These vertices are related to the kernel in eq.\ (\ref{bfkl})
by $C_i^2\sim K$.
As above, all lines in the ladder represent gluons.
The growth of the cross section results from the production
of many particles, whose
momenta are conveniently parameterized in components parallel
to the incoming momenta $k_a$ and $k_b$, and transverse components,
\begin{equation}
k_i=\alpha_ik_a+\beta_ik_b+k_{iT}\, .
\label{ksubi}
\end{equation}
The BFKL pomeron resums logarithms that result from configurations
in which the rapidities $y_i\sim \ln (\alpha_i/\beta_i)$, are
strongly ordered, but the transverse momenta are all of the
same order as the momentum transfer,
\begin{eqnarray}
\alpha_1 \gg \alpha_2 &\gg& \cdots \gg \alpha_n \nonumber \\
\beta_1 \ll \beta_2 &\ll& \cdots \ll \beta_n \nonumber \\
     k_{iT} &\sim& k_{jT}\, .
\label{strongorder}
\end{eqnarray}
It is the lack of ordering in transverse momentum that distinguishes
BFKL evolution.  Studies show
that kinematic signs of this evolution should
be present in final states \cite{bfklkinem}, and data from the
Tevatron and HERA are being closely scrutinized for evidence of
these effects.

\subsection{Soft and Hard Diffraction}

Ampitudes for soft and hard diffraction are illustrated in Fig.\ 5.  In 5a,
the lines in both the cut and uncut ladder are ordered in rapidity,
just as in Fig.\ 4.
\begin{figure}[ht]
\centerline{\epsffile{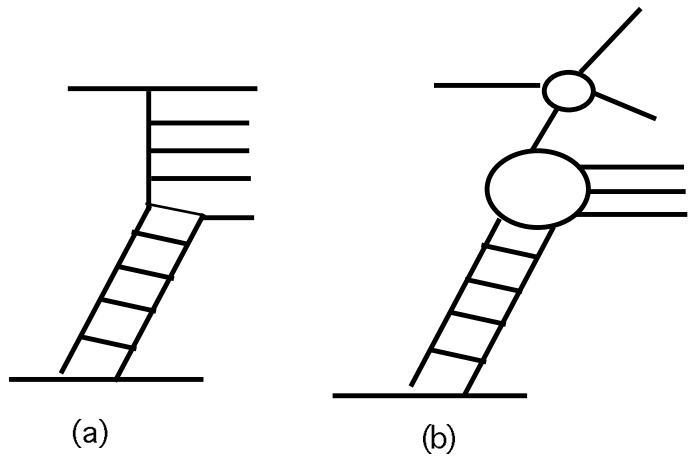}}
\caption{Schematic representations of soft (a) and hard (b)
diffraction.}
\label{fig5}
\end{figure}
Only the cut lines, closer in rapidity to the
incoming top line appear in the final state, however, leading to
a ``gap" in rapidity, and an apparent ``diffractive excitation"
of the top line.  The uncut ladder, a virtual pomeron, acts very
much like a virtual elementary particle.
Such events have been seen for some time, with
particles that emerge into the final state generally at low transverse
momentum.  More recently, however, events have been seen that have
both rapidity gaps and high momentum transfers
\cite{rapgap}.  A possible mechanism is
illustrated in Fig.\ 5b, in which the incoming top line undergoes
a hard scattering from a ``parton" of the pomeron.  This idea \cite{IngSch}
suggests that it might make sense to talk about the distributions
of partons in pomerons.  Such a formalism would be useful if
these cross sections enjoy the universality and incoherence
properties that underly normal factorization for inclusive
cross sections \cite{Collinspomphi}.  Beyond this, the concept
of rapidity gaps as a sign of color singlet exchange, even at
large momentum transfer \cite{singletrapgap} is an attractive concept.

As noted above, the BFKL formalism treats leading logarihms of
$s$.  For singlest exchange, these are single logarithms per loop,
$\alpha_s^n\ln^ns$.  More generally, elastic scattering
of, say, high-energy quarks may
involve both singlet and octet exchange
\begin{equation}
A = A_8T_8 + A_1 T_1\, ,
\label{coloramp}
\end{equation}
with the $T_i$ tensors in the color space.
Supposing that $t$ is large, but $|t|\ll s$, it is possible
to study the leading logarithms assoicated with soft
gluon exchange according to Sudakov evolution methods \cite{softSud}.
The results for the $A_i$, neglecting for simplicity the effects of the running
coupling, are of the schematic form \cite{K,StS}
\begin{equation}
A_8\sim e^{-\ln s\ln t}\, \, ,\, \, A_1 \sim e^{-\ln^2 t}\, .
\label{colorsud}
\end{equation}
This is consistent with BFKL resummation of logs of $s$ in
singlet exchange at fixed $t$.  For octet exchange, it is equivalent to
a ``Reggeized" gluon behavior \cite{BFKL}.
These results are also consistent with the observation of
rapidity gaps in hard scattering  at high energy, and
with their connection to singlet exchange.  They are also
relevant to a very different process, near-forward
hard elastic scattering \cite{LandDonn}, in which
an observed behavior of $t^{-8}$ may be
a sign of triple pomeron exchange between valence quarks \cite{StS}.

\subsection{Taming the Bare Pomeron}

As noted above, the bare pomeron violates the Froissart bound.
Considerable effort recently has gone into remedying this
defect, and to creating a formalism that reduces to the
BFKL analysis for moderate $s$, but which has built-in unitarity.
Here, I shall mention only a few examples.  Lipatov
has developed such a scheme, based on adding a subset
of nonleading logarithms in $s$ and employing an
expansion in $1/N$, with $N$ the number of colors.
In this scheme, the pomeron
is a sort of bound state of Reggeized gluons \cite{lipconf}.  Adding
more such gluons, the problem reduces to
finding the spectrum of a two-dimensional
field theory in impact parameter space.  Recently, Faddeev and
Korchemsky \cite{FadKor} have shown that the resulting model is
``solvable" in the sense of possessing infinite numbers of
conservation laws.  They do not, however, explicitly ``solve" it.

Another approach toward deciphering the BFKL pomeron
involves studying a model in which it arises self-consistently
in perturbation theory, the scattering of bound states of
heavy quarks \cite{qqbarbfkl}.  Although not yet realistic,
such models allows a quite detailed study of soft gluon
dynamics at large $N$ and leading logarithm.

\section{Resummation at the edge of phase space}

I now turn to applications of Sudakov resummation, beginning with
a typical example from event shapes, the thrust.

\subsection{Thrust}

The thrust in ${\rm e}^+{\rm e}^-$ annihilation is defined by
\begin{equation}
T= {\rm max}_{\hat n} {1\over Q} \sum_i\, |{\vec p}_i\cdot {\hat n}|\, ,
\label{thrust}
\end{equation}
with the sum over all particles
in the final state, and with a maximum taken over all unit vectors $\vec n$.
In perturbation theory, the fixed-thrust cross section
diverges at $T=1$, which is the elastic limit of two
lightlike back-to-back
jets.  A measure of this divergence is given at one loop by
\begin{equation}
{1\over \sigma_0}\,
\int_0^T dT' {d\sigma \over dT'} \sim 1- {\alpha_s\over \pi}C_F\; \ln^2(1-T)\,
{}.
\label{oneloopthrust}
\end{equation}
If the two jets have masses $p_i^2\sim (1/2)(1-x_i)Q^2$, then near
the edge of phase space $x_1=x_2=1$, we have
\begin{equation}
T \sim {1\over 2}\left [ (1-x_1)+(1-x_2)\right ]\, .
\label{twojetthrust}
\end{equation}
In the same region, however, the cross section
(simplified by neglecting
certain nonleading corrections) factors into the
product of the jet functions described above,
\begin{eqnarray}
{1\over \sigma_0}\,
{d\sigma \over dT'} &\sim& \int dx_1dx_2\; \delta\left(1-T-{1\over
2}(2-x_1-x_2)\right )
\nonumber \\
&\ & \quad\quad\quad \times
J_1((1-x_1)Q^2,Q)\; J_2((1-x_2)Q^2,Q)\, ,
\label{thrustfact}
\end{eqnarray}
each of which satisfies a Sudakov evolution equation, whose solution gives
\begin{eqnarray}
\int dx\, e^{-\nu(1-x)Q^2/2}\; &J&((1-x)Q^2,Q) \nonumber \\
&\sim&
\exp \left [ -\int_0^1 {du\over u}(1-e^{-u\nu}) \int_{u^2Q^2}^{uQ^2}
{d\mu^2\over \mu^2} \Gamma_J(\alpha_s(\mu^2))    \right ]\, .
\label{thrustresum}
\end{eqnarray}
This result \cite{thrustresum}, and its analog for the transverse momentum
distribution
in Drell-Yan and Z$_0$ production \cite{qtresum}, are very helpful in
understanding
high energy data.

\subsection{Threshold Resummation}

Another important application of Sudakov resummation is
to high-mass Drell-Yan, top and other cross sections
where the parton distributions are falling rapidly at
threshold.  The simplest example is the inclusive Drell-Yan cross
section,
\begin{equation}
{d\sigma \over dQ^2}
=
\sigma_0\, \sum_q\, \int_\tau^1 dz\;
\Phi_{q{\bar q}}(z)\;
\omega_{q{\bar q}}(z,\alpha_s(Q^2))\, ,
\label{DYfact}
\end{equation}
where the partonic flux is given by
\begin{equation}
\Phi_{q{\bar q}}(z)
= \int_0^1 {dx_adx_b\over x_ax_b}\; \delta\left(1-\tau/(zx_ax_b)\right)\;
Q_q^2\; \phi_q \left (x_a,Q^2\right )
\phi_{\bar q} \left (x_b,Q^2\right )\, ,
\label{Phitauz}
\end{equation}
with $Q_q$ the quark charge in units of electron charge and with
$\sigma_0=4\pi\alpha^2/3NQ^2s$ the Born cross section for $Q_q=1$.
Here the overall partonic invariant is ${\hat s}=Q^2/z$, with
threshold at $z=1$.

Specializing to DIS scheme, initial state radiation from the incoming
quark pair in Drell-Yan cancels against initial state radiation
in the DIS structure function, illustrated in Fig.\ 3 above.
The (infrared safe) radiation from the outgoing quark jet in DIS,
however, is not compensated.  But since, as indicated above,
the $x\rightarrow 1$ behavior of this jet can be computed, we
can give an explicit form for the hard-scattering cross section
$\omega_{q{\bar q}}$ in eq.\ (\ref{DYfact}) that includes all
$\ln^n(1-x)$ behavior in this limit.  Schematically, it is of the
form
\begin{equation}
\omega_{q{\bar q}}(z,\alpha_s(Q^2))
=
\delta(1-z)
-
\left [ {\rm e}^E(1-z,\alpha_s) F(1-z) \right ]_+\, ,
\label{omegaresum}
\end{equation}
where the function $F(1-z)$ is given by \cite{CoSt}
\begin{equation}
F(1-z)={1 \over \pi}\; \sin(\pi P_1)\Gamma(1+P_1)
+\dots\, ,
\label{Fdef}
\end{equation}
with $P_1$ the derivative of the exponent $E$ with respect
to $\ln (1-z)$,
\begin{equation}
P_1(1-z,\alpha_s)={d \over d\ln(1-z)}E(1-z,\alpha_s)\, .
\label{ponedef}
\end{equation}
$E(1-z,\alpha_s)$ is itself of the moment form,
\begin{equation}
E(n,\alpha_s)
=
\int_0^1 dx \left ({x^{n-1}-1 \over 1-x}\right )
\int_{(1-x)^2Q^2}^{(1-x)Q^2}
{d\mu^2 \over \mu^2}\, g_1(\alpha_s(\mu^2)) + \dots
\label{Edef}
\end{equation}
where $g_1$ is closely related to the jet anomalous dimension $\Gamma_J$,
discussed above,
\begin{equation}
g_1(\alpha_s)=2C_F\; {\alpha_s \over
\pi}+\dots=2\Gamma_J(\alpha_s)+\dots\, .
\label{geeone}
\end{equation}
Given the form of $E$, we expect the effect of these
corrections to be positive, and to raise the cross section
above low order predictions.  This expectation has been
borne out by explicit calculations in leading-logarithm
approximation for the DY process \cite{AMS}.

We have heard elsewhere at this conference that one-loop
QCD predictions fall short of the very high-$E_T$
jet cross section measured by CDF.  This cross section is
of the same general form as above, given schematically by
\begin{equation}
{ d\sigma_{1J} \over  dE_T}\bigg |_{y=0}
=
\int_{2E_T/\sqrt{s}}^1\;
\Phi(z)\; { d{\hat \sigma}_{1J} \over dE_T}
({\hat s}=zs)\; dz\, .
\label{jetsigma}
\end{equation}
It is natural to ask whether this effect might be
due to resummable soft gluon effects.  Now we must note that
in jet production final-state as well as
initial state interactions may be important.  Nevertheless,
final-state interactions cancel in any cross section with
finite angular and energy resolutions, leaving the same,
universal initial-state corrections as in DY.  It will
be of great interest to test these ideas in the
coming months, since an excess of high-energy
jets relative to QCD predictions could indicate
profound new physics.

\subsection{Regulated Resummation}

Eq.\ (\ref{Edef}) for the resummed exponent has a problem
at small $1-x$, where the perturbative running coupling may diverge.
To make sense of (\ref{Edef}), one may use a cutoff in $x$
or $k_T^2$.  Another possibility, following the example of
ref.\ \cite{Mu85}, is to define the integral as a principal value
\cite{CoSt2}.  There is no special status for such a definition,
but it enables us to express $E$ explicitly in terms of special
functions, to be specific as a series of exponential integrals.
In fact, any such definition of the perturbative expansion can
be made only at the price of adding new nonperturbative
parameters to the theory, as
\begin{equation}
E_{NP} = \sum_{n\ge 1} {A_n \over Q^n}\, .
\label{enp}
\end{equation}
Cutoff integrals are probably numerically simpler.  A sample
application to resummed leading logarithms was studied in
\cite{AMS} for the Drell-Yan process,
\begin{equation}
\int_0^1 dx\; {x^{n-1} -1 \over 1-x}
\rightarrow
\int_0^{1-\lambda/\Lambda} dx\; {x^{n-1} -1 \over 1-x}\, .
\label{cutoff}
\end{equation}
Order by order, dependence on $\lambda$ is higher twist.
For large enough
$Q^2$, a cutoff resummation of
this form is relatively insensitive to $\lambda$, but
grows with $\tau$ at fixed $Q^2$ \cite{AMS}.  Studies of the behavior at
fixed $s$ are currently underway.

Finally, we may note that a cutoff prescription was
employed in \cite{LSvN} to define resummed perturbation
theory for top production, with partonic cross sections
of the form
\begin{equation}
\sigma_{ij}(z,m_t^2,s_0)
=
\int_{s_0}^{s-2m_t\sqrt{s}}
ds_4\; e^{{\bar E}(s_4)}\; {d\sigma^{(0)}_{ij} \over ds_4}\, ,
\label{lsvn}
\end{equation}
where $s_4=0$ at threshold, where ${\bar E}(s_4)$ is a
resummed exponent related to (\ref{Edef}) and where
$\sigma^{(0)}_{ij}$ is the Born cross section.

\section{Infrared Renormalons and a New Source of Power Corrections}

Expanding the
running coupling  integrand in eq.\ (\ref{Edef})
in terms of $\alpha_s(Q^2)$, we derive integrals of the form
\begin{equation}
(-b_2)^m\alpha_s^m\; \int_0^{Q^2} {dk_T^2\over Q^2}\;\ln^m
\left({k_T^2\over Q^2}\right )
=b_2^m\alpha_s^m\; m!\, ,
\label{irrenorm}
\end{equation}
in which the singularity in the running coupling is
reflected in factorial growth of expansion coefficients
in perturbation theory.  This phenomenon is commonly
referred to as an ``infrared renormalon" \cite{tHooft}.
A closer analysis shows that such behavior may be
interpreted as an ambiguity in the perturbative
expansion, which may be removed by modifying the
expansion to make it convergent, while at the same
time adding a new term of the form of eq.\ (\ref{enp}),
$A_N/Q^N$, with
$Q$ the momentum transfer.  The classic case \cite{Mu85}
is the total cross section in ${\rm e}^+{\rm e}^-$ annihilation,
where the nonperturbative term is of the form
$A_4/Q^4$, with $A_4=\langle F^2\rangle_0$, the
vacuum expectation value of the squared gluon field strength.
A similar analysis for resummed jet shapes \cite{StArg} and
Drell-Yan cross sections \cite{CoSt2} shows that their
resummed exponents potentially have a much stronger nonperturbative
behavior \cite{W,KS},
\begin{equation}
E=E_{\rm PT}^{({\rm reg})}+{\Lambda \over \delta Q}A\, ,
\label{interp}
\end{equation}
where $\delta=0$ in the elastic limit (infinitely
collimated jets, for instance).
As in the case of the total ${\rm e}^+{\rm e}^-$
cross section, it is possible to find a field-theoretic
analog to $A$, but in this case it is in terms of
a field-strength, integrated over the classical
paths of the
relevant partons.  For example,
for a two-jet cross section, with directions are $p_1$ and $p_2$,
we find \cite{KS}
\begin{eqnarray}
\Lambda A &=& \langle 0|{\bar {\rm T}}
\left ( \Phi^\dagger_{p_1}(0,\infty)
({\cal F}_{0p_1}^\dagger(0) -{\cal F}_{0p_2}(0,\infty))
\Phi_{p_2}(0,\infty)
\right ) \nonumber \\
&\ & \quad\quad\quad  \times {\rm T} \left (\Phi_{p_1}^\dagger(0,\infty)
\Phi_{p_2}(0,\infty)
\right ) |0\rangle\, .
\label{Aop}
\end{eqnarray}
This matrix element is gauge invariant, with the
$\Phi$'s defined in terms of ``Wilson lines",
and the operator $\cal F$ in terms of the
field strength.  The rather  general form of the operator suggests
that these new corrections, suppressed by a single
power of the large momentum scale, may possess
universality properties.  The concept of universality
in this context has already been given much study
\cite{KS,DokWeb},
and its promise and limitations have been
discussed in \cite{KSMord}.

\section{Conclusions}

Soft-gluon resummation allows varied applications of quantum field
theory that are relevant to QCD at current energies.  One goal of this
program is to achieve a new level of precision by combining large
corrections to all orders.  It also affords potential insights into
features of nonperturbative structure, through the high-order
behavior of the perturbation series.

Regarding specific applications, resummed threshold corrections to
very high energy jet cross sections urgently need more study,
especially because they are relevant to signs of new physics.  Also,
studies of $-t\ll s$ hard elastic and jet cross sections are attractive
for their relevance to hard pomeron and Sudakov
effects and to the valence-quark structure  of the nucleon.  Finally,
infrared renormalons from resummation
imply the existence a new class of
nonperturbative parameters that may be measured in, for instance,
jet shape analysis, and which may shed new light on the
perturbative/nonperturbative interface in QCD.
\smallskip

It is a pleasure to thank Lyndon Alvero,
Harry Contopanagos, Gregory Korchemsky and Jack Smith for many
helpful conversations and explanations.
This work was supported in part by the National Science Foundation under
grant PHY9309888.

%

%


\end{document}